\documentclass[conference]{IEEEtran}

\usepackage{cite}      

\usepackage{graphicx}  

\usepackage{amsmath}    
\begin{document}

\title{Applications of Feedback Control in Quantum Systems}

\author{\authorblockN{Kurt Jacobs}
\authorblockA{Quantum Science and Technologies Group, Hearne Institute for Theoretical Physics,\\
Louisiana State University, 202 Nicholson Hall, Tower Drive, Baton Rouge, LA 70803, USA}
}

\maketitle

\begin{abstract}
We give an introduction to feedback control in quantum systems, as well 
as an overview of the variety of applications which have been explored to date. This 
introductory review is aimed primarily at control theorists unfamiliar with quantum 
mechanics, but should also be useful to quantum physicists interested in applications 
of feedback control. We explain how feedback in quantum systems differs from that 
in traditional classical systems, and how in certain cases the results from modern 
optimal control theory can be applied directly to quantum systems. In addition to noise 
reduction and stabilization, an important application  
of feedback in quantum systems is adaptive measurement, and we discuss the various 
applications of adaptive measurements.  We finish by describing 
specific examples of the application of feedback control to cooling and 
state-preparation in nano-electro-mechanical systems and single trapped atoms. 
\end{abstract}

\section{Introduction}

\begin{figure}
\centering
\includegraphics[width=3.0in]{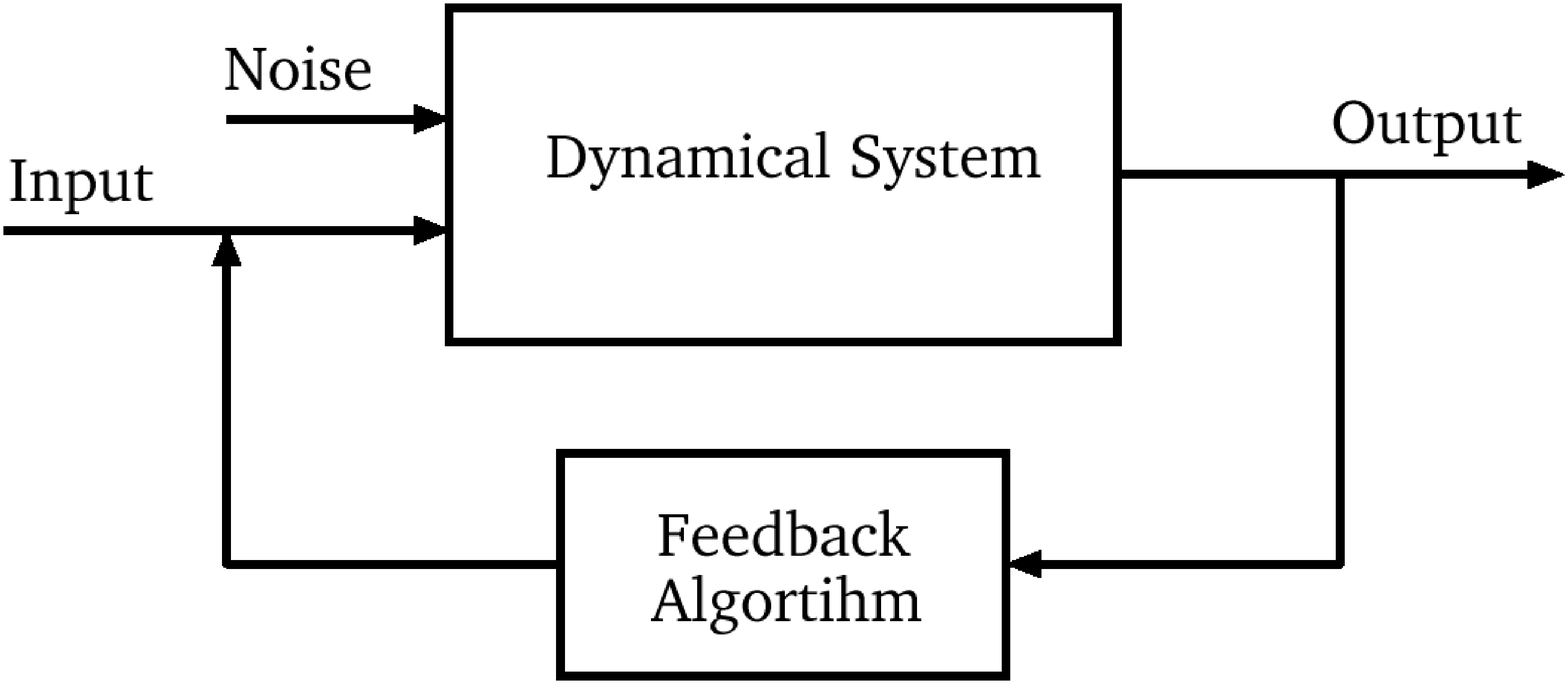} 
\caption{A diagrammatic depiction of the process of feedback control. 
The output of the noisy dynamical system is monitored. The resulting
measured signal is processed by a device which calculates the required
input (the feedback) as a functional of this signal. The precise
functional used is called the {\em feedback algorithm}. The goal of
the feedback is often to minimize the effect of the noise on the
dynamics of the system, but it can also be to modify the behavior of
the system in some other way.}
\label{fig1}
\end{figure}

While most readers will be familiar with the notion of {\em feedback
control}, for completeness we begin by defining this term. Feedback
control is the process of monitoring a physical system, and using this
information as it is being obtained (in real time) to apply forces to
the system so as to control its dynamics. This process, which is
depicted in Figure~\ref{fig1}, is useful if, for example, the system
is subject to noise.

Since quantum mechanical systems, including those which are
continually observed, are dynamical systems, in a broad sense the
theory of feedback control developed for classical dynamical systems
applies directly to quantum systems\footnote{Here we use the term {\em
classical} to refer to systems which are the traditional purview of
control theory - mechanical systems obeying Newton's equations, and
electrical systems obeying Maxwell's equations}. However, there are
two important caveats to this statement.  The first is that most of
the exact results which the theory of feedback control provides,
especially those regarding the optimality and robustness of control
algorithms, apply only to special subclasses of dynamical systems. In
particular, most apply to linear systems driven by Gaussian
noise~\cite{Kseq,Whittle}.  Since observed quantum systems in general
obey a non-linear dynamics\footnote{The dynamics of an unobserved
quantum system is given by Schr\"odinger's equation, which is
linear. However, the act of continually observing a quantum system 
will in general induce a
non-linear dynamics.}, an important question that arises is whether
exact results regarding optimal control algorithms can be derived for
special classes of quantum systems.

In addition to the need to derive results regarding optimality which
are specific to classes of quantum systems, there is a property that
sets feedback control in quantum systems apart from that in other
systems. This is the fact that in general the act of measuring a
quantum system will alter it. That is, measurement induces dynamics in
a quantum system, and this dynamics is noisy as a result of the
randomness of the measurement results. Thus, when considering the
design of feedback control algorithms for quantum systems, the design
of the algorithm is not independent of the measurement process. In
general different ways of measuring the system will introduce
different amounts of noise, so that the search for an optimal feedback
algorithm must involve an optimization over the manner of measurement.

In what follows we will discuss a number of explicit examples of
feedback control in a variety of quantum systems, and this will allow
us to give specific examples of the dynamics induced by
measurement. Before we examine such examples however, it is worth
presenting the general equations which describe feedback control in
quantum systems, in analogy to those for classical systems. In
classical systems, the state-of-knowledge of someone observing the
system is given by a probability density over the dynamical variables
(a phase-space probability density). Let us consider for simplicity a
single particle, whose dynamical variables are its position, $x$ and
momentum, $p$. If the observer is continually monitoring the position
of the particle, then her stream of measurement results, $r(t)$ is
usually described well by
\begin{equation}
    dr  = x(t) dt + \frac{dV(t)}{\sqrt{\gamma}} \, ,
\end{equation}
where in each time interval $dt$, $dV(t)$ is a Gaussian random
variable with variance $dt$ and a mean of zero. Such a Gaussian noise
process is called Wiener noise~\footnote{Accessible introductions to Wiener 
noise are given in~\cite{WienerIntroPaper,KJPhD}}.  The constant $\gamma$ 
determines the
relative size of the noise, and thus also the {\em rate} at which the
measurement extracts information about $x$; when $\gamma$ is
increased, the noise decreases, and it therefore takes the observer
less time to obtain an accurate measurement of $x$.

As the observer obtains information, her state-of-knowledge regarding the system, $P(x,p)$,
evolves. The evolution is given by the Kushner-Stratonovich (K-S) Equation. This is
\begin{equation}
   dP = \left[ -\frac{p}{m}\partial_x - F(x,t)\partial_p\right] P dt + \sqrt{\gamma} (x - \langle x(t) \rangle) P dW ,
\end{equation}
where $m$ is the mass of the particle, $F(x,t)$ is the force on the particle, 
\begin{equation}
   \langle x(t) \rangle = \int_{-\infty}^{\infty} \int_{-\infty}^{\infty} \!\!\!\! x \, P(x,p) \; dx dp 
\end{equation}
is the expectation value of $x$ at time $t$, and 
\begin{equation}
  dW(t) = \sqrt{\gamma}(x - \langle x \rangle) dt + dV(t) 
\end{equation}
turns out to be a Wiener noise, uncorrelated with the probability density
$P$.  Because of this we can alternatively write the stream of measurement results as 
\begin{equation}
   dr = \langle x(t) \rangle dt + \frac{dW}{\sqrt{\gamma}} \, .
\end{equation}
The above will no doubt be familiar to the majority of the readership. For linear 
dynamical systems the K-S equation reduces to the equations of the well-known 
Kalman-Bucy filter~\cite{Kseq}. 

The K-S equation is the essential tool for describing feedback control; it tells us what 
the observer knows about the system at each point in time, and thus the information 
that he or she can use to determine the feedback forces at each point in time. In addition, 
when we include these forces in the system dynamics, the resulting K-S equation, in telling 
us the observer's state-of-knowledge is also telling us how effective is our feedback control:
the variance of this state-of-knowledge, and the fluctuations of its mean (note that these are 
two separate things) tell us the remaining uncertainty in the system. The K-S equation thus allows us to 
design and evaluate feedback algorithms. 

The description of dynamics and continuous measurement in quantum
mechanics is closely analogous to the classical case described above. In quantum
mechanics, however, the observer's state-of-knowledge must be
represented by a matrix, rather than a probability density. This
matrix is called the {\em density matrix}, and usually denoted by
$\rho(t)$. The dynamical variables are also represented by
matrices. If the position is represented by the matrix $X$, then the
expectation value of the particle's position at time $t$ is given by
$\langle x (t) \rangle = \mbox{Tr}[\rho(t) X]$~\cite{Sakuri}. While
the notion that a state-of-knowledge is described by a matrix will
appear very strange to most of the readership, don't let this put you
off --- when we consider feedback control we will always discuss it in
terms of standard physical quantities such as the expectation values,
variances, or probability densities for the dynamical variables.  The
reason we speak of the density matrix as representing the observer's
state-of-knowledge is because all these quantities can be obtained
directly from the density matrix.

The dynamics of an unobserved quantum system may be written as $i\hbar
\dot{\rho} = [H,\rho] \equiv H\rho - \rho H$ for a given matrix $H$
called the Hamiltonian ($\hbar$ is Planck's constant). If an observer 
makes a continuous measurement
of a particle's position, then the full dynamics of the observer's
state-of-knowledge is given by the quantum equivalent of the
Kushner-Stratonovich equation. This is
\begin{eqnarray}
d\rho &=& - \frac{i}{\hbar} [H,\rho]dt - (\gamma/8) [X, [X, \rho]]dt \nonumber\\
      && + \frac{\sqrt{\gamma}}{2} ( (X - \langle x(t) \rangle)\rho + \rho (X - \langle x(t) \rangle) ) dW ,
\label{sme}
\end{eqnarray}
where the observer's stream of measurement results is\footnote{The
stream of measurement results is usually referred to as the {\em
measurement record}.}
\begin{equation}
   dr = \langle x(t) \rangle dt + \frac{dW}{\sqrt{\gamma}} .
\end{equation}
This is referred to as the {\em Stochastic Master Equation} (SME), and
was first derived by Belavkin~\cite{BelavkinLQG}. This is very similar
to the K-S equation, but has the extra term $[X,[X,\rho]]$ that
describes the noisy dynamics (or {\em quantum back-action}) which is
introduced by the measurement.  In the quantum mechanics literature the 
measurement rate $\gamma$ is often referred to as the {\em measurement 
strength}. The SME is usually derived directly from quantum measurement
theory~\cite{WM93,Brun02} without using the mathematical machinery of
filtering theory. A recent derivation for people familiar with
filtering theory may be found in~\cite{Bouten06}. Armed with the
quantum equivalent of the K-S equation, we can proceed to consider
feedback control in quantum systems\footnote{A further discussion of
analogies between quantum and classical descriptions of
state-estimation and feedback is given in references~\cite{DJ,DHJMT}}.

Interpreting ``applications of feedback control in quantum systems''
in a broad sense, it would appear that one can break most such
applications into three general classes. While these classes may be
somewhat artificial, they are a useful pedagogical tool, and we will
focus on specific examples from each group in turn in the following three
sections.  The first group is the application of results and techniques from
classical control theory to the general theory of the control of
quantum systems. This includes the application control theory to
obtain optimal control algorithms for special classes of quantum
systems.  An example of this is the realization that classical LQG
control theory can be applied directly to obtain optimal control
algorithms for observed linear quantum
systems~\cite{BelavkinLQG,Yanagisawa98,DJ}. We will discuss this and
other examples in Section~\ref{Systems}.

It is worth noting at this point that while the direct application of
classical control theory to quantum systems is very useful, it is not
the only approach to understanding the design of feedback
algorithms in quantum systems. Another approach is to try to gain
insights into the relationship between measurement (information
extraction) and disturbance in quantum mechanics which are relevant to
feedback control. Such questions are also of fundamental interest to
quantum theorists since they help to elucidate the information
theoretic structure of quantum mechanics. References~\cite{FJ}
and~\cite{DJJ} take this approach, elucidating the 
information-disturbance trade-off relations for a two-state system
(the simplest non-linear quantum system), and exploring the effects of
this on the design of feedback algorithms.

The second group is the application of feedback control to classes of
control {\em problems} which arise in quantum systems, some of which
are analogous to those in classical systems, and some of which are
peculiar to quantum systems. A primary example of this is in adaptive
measurement, where feedback control is used during the measurement
process to change the properties of the measurement. This is usually
for the purpose of increasing the information which the measurement
obtains about specific quantities, or increasing the rate at which
information is obtained. We will discuss such applications in
Section~\ref{Problems}.

The third group is the design and application of feedback algorithms
to control specific quantum systems.  Examples are applications to the
cooling of a nano-mechanical resonator and the cooling of a single atom 
trapped in an optical cavity. We will discuss these in Section~\ref{Physical}.

%

\section{Optimal Control for Linear Quantum Systems}
\label{Systems}
Most readers of this article will certainly be familiar with the
classical theory of optimal Linear-Quadratic-Gaussian (LQG)
control. This provides optimal feedback algorithms for linear systems
driven by Gaussian noise, and in which the observer monitors some
linear combination of the dynamical variables. In LQG control, the
control objective is the minimization of a quadratic function of the
dynamical variables (such as the energy). It turns out that for a
restricted class of observed quantum systems (those which are linear
in a sense to be defined below), this optimal control theory can be
applied directly. This was first realized by Belavkin
in~\cite{BelavkinLQG}, and later independently by Yanagisawa and
Kimura~\cite{Yanagisawa98} and Doherty and Jacobs~\cite{DJ}.

Quantum mechanical systems whose Hamiltonians are no more than
quadratic in the dynamical variables are referred to as linear quantum
systems, since the equations of motion for the matrices representing
the dynamical variables are linear. Further, these linear equations
are precisely the same as those for a classical system subject to the
same forces. The simplest example is the Harmonic oscillator. If we
denote the matrices for position and momentum as $X$ and $P$
respectively, then the Hamiltonian is
\begin{equation}
  H = \frac{P^2}{2m} + \frac{1}{2}m\omega^2 X^2
\end{equation}
where $m$ is the mass of the particle, and $\omega$ is the angular 
frequency with which it oscillates. The resulting equations for $X$ and $P$ are 
\begin{eqnarray}
   \dot{X} & = & P/m \\
   \dot{P} & = & -m\omega X ,
\end{eqnarray}
which are of course identical to the classical equations for the
dynamical variables $x$ and $p$ in a classical harmonic
oscillator. Further, it turns out that if an observer makes a
continuous measurement of any linear combination of the position and
momentum, then the SME for the observer's state of knowledge of the
quantum system reduces to the Kushner-Stratonovich equation, which in
this case, because the system is linear, is simply the Kalman-Bucy
Filter Equations\footnote{Strictly speaking, for this to be true the
initial state of the system must be a Gaussian probability density in
phase space. However, if this is not the case, the dynamics induced by
the measurement is such that the density will become Gaussian over
time. Thus after a sufficient time the SME will approximately reduce
to the Kalman-Bucy Filter for any initial state.}.

However, there is one twist. The Kalman-Bucy equations one obtains for
the quantum system are those for a classical harmonic oscillator
driven by Gaussian noise of strength $\hbar\gamma$.  This comes from
the extra term in the SME which describes the ``quantum back-action''
noise generated by the measurement.  Since the observer's state of
knowledge evolves in precisely the same way as for the equivalent
linear classical system, albeit driven by noise, we can apply
classical LQG theory to these quantum systems. If we apply linear
feedback forces, then, for a fixed measurement strength $\gamma$, the
quantum mechanics will tell us how much noise the system is subject
to, and LQG theory will tell us the resulting optimal feedback
algorithm for a given quadratic control
objective~\cite{BelavkinLQG,DJ}.

Note however, that since the noise driving the system depends upon the
strength of the measurement, then the performance of the feedback
algorithm will also depend upon the strength of the
measurement. Moreover, the performance of the algorithm will be
influenced by two competing effects: as the measurement gets stronger,
we can expect the algorithm to do better as a result of the fact that
the observer is more rapidly obtaining information. However, as the
measurement gets stronger, the induced noise also increases, which
will reduce the effectiveness of the algorithm. We can therefore expect
that there will be an optimal measurement strength at which the feedback
is most effective. In a linear quantum system one therefore must first
find the optimal feedback algorithm using LQG control theory, and then
perform a second optimization over the measurement strength. This is
not the case in classical control. An explicit example of optimizing
measurement strength may be found in~\cite{DW04}.

The application of LQG theory to linear quantum systems will be useful
when we examine the control of a nanomechanical resonator in
Section~\ref{Physical}.  The close connection between linear quantum
and classical systems allows one to apply other results from classical
control theory for linear systems to quantum linear systems. Transfer
function techniques have been applied to linear quantum systems by
Yanagisawa and Kimura~\cite{Yanagisawa03a,Yanagisawa03b}, and D'Helon
and James have elucidated how the small gain theorem can be applied to
linear quantum optical networks~\cite{DHelon05x}. Finally, it is
possible to obtain exact results for the control of linear quantum
systems for at least one case beyond LQG theory: James has extended
the theory of risk-sensitive control to linear quantum
systems~\cite{James04}.

For nonlinear quantum systems, naturally many of the approaches
developed for classical non-linear systems can be expected to be
useful. A few specific applications of methods developed for classical
nonlinear systems have been explored to date. One example is the use
of linearization to obtain control algorithms~\cite{DHJMT}, and
another is the application of a classical guidance algorithm to the
control of a quantum system~\cite{Ralph04}. A third example is the
application of the projection filter technique to obtain approximate
filters for continuous state-estimation of nonlinear quantum optical
systems~\cite{Handel05}. The Bellman equation has also been
investigated for a two-state quantum system in~\cite{BEB}; it is not
possible to obtain a general analytic solution to this equation for
such a system, and as yet no-one has attempted to solve this problem
numerically. As quantum systems become increasingly important in the
development of technologies, no doubt many more techniques and results
from non-linear control theory will be applied in such systems.

\section{Adaptive Measurement}
\label{Problems}
The objective of LQG control is to use feedback to minimize some
quadratic function of the dynamical variables, and this is natural if
one wishes to maintain a desired behavior in the presence of
noise. While stabilization and noise reduction in dynamical systems is
a very important application of feedback control, it is by no means
the only application. Another important class of applications is {\em
adaptive measurement}.

An adaptive measurement is one in which the measurement is altered as
information is obtained. That is, a process of feedback is used to
alter the measurement as it proceeds, rather than altering the
system\footnote{In fact, from the point of view of the measurement
alone, adjusting the measurement is always equivalent to adjusting the
state of the system.} The primary distinction between adaptive
measurement and more traditional control objectives however is that
the goal of the former is usually to optimize some property of the
information obtained in the measurement process rather than to control
the dynamics of the system.

Adaptive measurement has, even at this relatively early stage in the
development of the field of quantum feedback control, found many
potential applications in quantum systems. The reason for this is due
to the interplay of the following things. The first is that unlike
classical states the majority of quantum states are not fully
distinguishable from each other, even in theory, and only carefully
chosen measurements will optimally distinguish between a given set of
states. The second is that because quantum measurements generally
disturb the system being measured, one must also choose one's
measurements very carefully in order to extract the maximal
information about a given quantity (lest the measurement disturb this
quantity). Combining these two things with the fact that one is
usually limited in the kinds of measurements one can perform, due to
the available physical interactions between a given system and
measuring devices\footnote{In optical systems, for example, ultimately
all one can do is to count photons, and one must therefore construct
ways to measure optical phase indirectly.}, it is frequently
impossible to implement optimal measurements in quantum systems. The
use of adaptive measurement increases the range of possible
measurements one can make in a given physical situation, and in some
cases allows optimal measurements to be constructed where they could
not be otherwise.

\subsection{State-Discrimination and Parameter Estimation}
As far as the author is aware, the first application of quantum
adaptive measurement was introduced by Dolinar in
1973~\cite{Dolinar73,Geremia04}. Here the problem involves
communicating with a laser beam, where each bit is encoded by the presence
or absence of a pulse of laser light.  Quantum effects become
important when the average number of photons in each pulse is small
(e.g. $\leq 10$). In fact, it is not possible to completely distinguish between 
the presence or absence of a pulse. The reason for this is that the 
quantum nature of the pulse of laser light is such that there is always a finite probability
that there are no photons in the pulse. It turns out that the optimal
way of distinguishing the two states is by mixing the pulse with
another laser beam at a beam splitter, and detecting the resulting
combined beam.  In this case both input states will produce photon
clicks. The optimal procedure is to vary the amplitude of the mixing
beam with time, and in particular to use a process of feedback to
change this amplitude after the detection of each photon. This
feedback procedure distinguishes the states maximally well within the
limits imposed by quantum mechanics (referred to as the Helstrom
bound~\cite{Helstrom}).

The objective of Dolinar's adaptive measurement scheme is to
discriminate maximally {\em well} between two states. Alternatively
one may wish to discriminate maximally {\em fast}. To put this another
way, one may wish to maximize the amount of information which is
obtained in a specified time, even if it is not possible to obtain all
the information in that time. Such considerations can be potentially
useful in optimizing information transmission rates when the time taken to prepare
the states is significant. In~\cite{Jacobs06x} it is shown that
adaptive measurement can be used to increase the speed of
state-discrimination. Of particular interest in quantum control theory
are situations which reveal differences between measurements on
quantum systems and those on their classical counterparts. The 
rapid-discrimination adaptive measurement scheme of~\cite{Jacobs06x} 
is one such example. The reason for this
is that in the case considered in~\cite{Jacobs06x} it is only possible
to use an adaptive algorithm to increase the speed of discrimination
if quantum mechanics forbids perfect discrimination between them. 
Since all classical states are
completely distinguishable (at least in theory), this adaptive
measurement is only applicable to quantum systems. The question of
whether or not there are more general situations which provide
classical analogues of this adaptive measurement is an open question,
however.

The problem of quantum state-discrimination is a special case of {\em
parameter estimation}. In parameter estimation, the possible states that 
a system could have been prepared in (or alternatively the possible 
Hamiltonians that may describe the dynamics of a system) are 
parametrized in some way. The observer then tries to determine the 
value of the
parameter by measuring the system. 
When discriminating two states the parameter has only two discrete 
values: in the above case it is the amplitude of
the laser pulse, which is either zero or non-zero. In a more general 
case one wishes to determine a continuous parameter. An example 
of this is the detection of a force. In
this case a simple system which feels the force, such as a quantum
harmonic oscillator, is monitored, and the force is determined from
the observed dynamics. Two examples of this are the atomic force
microscope (AFM)~\cite{Milburn94} and the detection of gravitational
waves~\cite{Braginsky03}. Adaptive measurement is useful in parameter
estimation because measurements which are not precisely tailored will
disturb the system so as to degrade information about the
parameter. To the author's knowledge no-one has yet investigated
adaptive measurement in force estimation, although the subject is
discussed briefly in~\cite{Verstraete01}. However, the use of adaptive
measurement for the estimation of a magnetic field using a cold atomic 
cloud has been investigated in~\cite{Stockton04,Geremia05}, and for 
the estimation of a phase shift imparted on a light beam has
been explored in references~\cite{Berry01,Berry02,Pope04}. 
In~\cite{Pope04} the authors show that adaptive measurement 
outperforms other known kinds of measurements for estimation of 
phase shifts on continuous beams of light.

\subsection{Constructing New Measurements}
It turns out that there are certain kinds of measurements which it is
very difficult to make because the necessary interactions between the
measurement device and the system are not easily engineered. One such
example in quantum mechanics is a measurement of what is referred to
as the ``Pegg-Barnett'' or ``canonical'' phase~\cite{Pegg88}. There are
subtleties in defining what one means by the phase of a quantum
mechanical light beam (or, more precisely, in obtaining a definition
with all the desired properties). Astonishingly, the question was not
resolved until 1988, when Pegg and Barnett constructed a definition
which has the desired properties for all practical purposes. This is
called the {\em canonical phase} of a light beam.

Now, the most practical method of measuring light is to use a photon
counter.  However, it turns out that it is not possible to use a
photon counter, even indirectly, to make a measurement which measures
precisely canonical phase\footnote{Strictly speaking, it is not
possible to make a {\em deterministic} measurement of canonical phase
using a photon counter. It is possible to make a measurement that
succeeds with some non-unity probability, a fact shown
in~\cite{Pregnell06}.} Nevetheless, in~\cite{Wiseman95} (see
also~\cite{Wiseman97,Wiseman98}) Wiseman showed that the use of an
adaptive measurement process allows one to more closely approximate a
canonical phase measurement.  For a light pulse which has at most one
photon, this adaptive phase measurement measures precisely canonical
phase. Wiseman's adaptive phase measurement has now been realized
experimentally~\cite{Armen02}.

\subsection{Rapid State-Preparation}
One application of feedback control is in preparing quantum systems in
well-defined states.  Due to noise from the environment, the state of
quantum systems which have been left to their own devices for an
appreciable time will contain considerable uncertainty. Many quantum
devices, for example those proposed for information
processing~\cite{Spiller05}, require that the quantum system be
prepared in a state which is specified to high precision.  Naturally
one can prepare such states by using measurement followed by control -
that is, using a measurement to determine the state to high precision,
and then applying a control field to move the system to the desired
point in phase space.

Since the rate at which a given measuring device will extract
information is always finite, one can ask whether it is possible to
increase the rate at which information is extracted by using a process
of adaptive measurement. That is, to adaptively change the measurement 
as the observer's state-of-knowledge changes, so that the uncertainty
(e.g. the entropy) of the observer's state-of-knowledge reduces 
faster. We will refer to the process of reducing the entropy 
as {\em purifying} the qubit (a term taken from the jargon of quantum 
mechanics). It turns out that it is indeed possible to increase
the rate of purification using adaptive measurement. 

For a two-state quantum system (often referred to as a {\em qubit}), an 
adaptive measurement is presented in~\cite{rapidP} that will speed-up 
the rate of purification in a certain sense. Specifically, it will increase the 
rate of reduction of the {\em average entropy} of the qubit
(where the average is taken over the possible realizations of the
measurement - e.g. over measurements on many identical qubits) by a factor of two. 
This adaptive measurement has two further interesting properties. The 
first is that there is no analogous adaptive scheme for the equivalent 
measurement on a classic two-state system (a single bit); the speed-up 
in the reduction of the average entropy is a quantum mechanical effect. 

The second property has to do with the statistics of the entropy
reduction. For a fixed measurement, while the entropy decreases with
time {\em on average}, on any given realization of the measurement the entropy
fluctuates randomly as measurement proceeds. For the adaptive 
measurement however, in the limit of strong feedback, the entropy reduces {\em
deterministically}. For a finite feedback force, there will always
remain some residual stochasticity in the entropy reduction, but this
will be reduced over that for the fixed measurement~\cite{rapidP,KJSPIE04}.
Thus if one is preparing many qubits in parallel, this adaptive measurement 
will reduce the spread in the time it takes the qubits to be prepared.  

The above adaptive algorithm does not speed up purification in every sense 
of the word, however: Wiseman and Ralph~\cite{Wiseman06x} have recently 
shown that while the above adaptive measurement reduces the {\em average 
entropy} of an ensemble of qubits more quickly, quite 
surprisingly it actually {\em increases} the {\em average time} it takes 
to prepare a given value of the entropy by a factor of two! In this sense, therefore, 
the above measurement strategy does not, in fact, speed up preparation. The 
reason for this is that when one considers an ensemble of spins, the majority purify 
quickly, whereas the average value of the entropy across the ensemble is 
increased considerably by a small number of straggling qubits that purify 
slowly. Thus, the adaptive measurement works by decreasing the time taken 
for the stragglers, but {\em increasing} the time taken for the majority, so that, 
for strong feedback, all qubits take the same time to reach a given 
purity.  Thus the feedback algorithm constructed in~\cite{rapidP} does not 
speed up the average time a given qubit will take to reach a target entropy, 
and this will be the important quantity if one is preparing qubits in sequence. 
Application of the above results to rapid purification of superconducting 
qubits is analyzed in~\cite{Ralph06,Griffith06x}. 

Is it possible, therefore, to use adaptive measurement to speed up the 
average preparation time? While the answer for a single qubit is almost 
certainly no, the answer is almost certainly yes in general: 
In~\cite{Combes06} the authors show that, for a quantum system with $N$ 
states it is possible to speed up the rate at which the average entropy is 
reduced by a factor proportional to $N$. While it is not shown directly that this 
algorithm also decreases the {\em average time} to reach a given target 
purity, it is fairly clear that this will be the case, although more work remains 
to be done before all the answers are in.  

Another application of feedback control involving purification of states has been 
explored in~\cite{SHM,vanHandel05}.  In this case, rather than the rate 
of purification, the authors are concerned with using feedback control during 
the measurement to obtain a specific final state, and in particular when the 
control fields are restricted. 

In this section we have been considering the application of feedback in quantum 
systems to problems which lie outside the traditional applications of noise reduction 
and stabilization. Most of these fall under the category of adaptive measurement, 
and these we have discussed above. One that does not is quantum 
error correction. The goal of such a process is noise reduction, but with the 
twist that the state of the system must be encoded in such a way that the 
controller does not disturb the information in the system. While we will not discuss this 
further here, the interested reader is referred 
to~\cite{Ahn02,Sarovar04,Ahn04} and references therein.

\section{Controlling Nanoscopic Systems} 
\label{Physical}

We now present examples of feedback control applied to two specific quantum systems. The first is a nano-mechanical resonator, and the second is a single atom trapped in an optical cavity. In both cases the goal of the feedback algorithm will be to reduce the entropy of the system and prepare it in its ground state. A nano-mechanical resonator is a thin, fairly ridged bridge, perhaps 200 nm wide and a few microns long. Such a bridge is formed on a layered wafer by etching out the layer beneath. If one places a conducting strip along the bridge and passes a current through it, the bridge can be made to vibrate like a guitar string by driving it with a magnetic field. So long as the amplitude of the oscillation is relatively small, the dynamics is essentially that of a harmonic oscillator~\cite{Blencowe04}. One of the primary goals of research in this area is to observe quantum behavior in these oscillators~\cite{Cho03}. The first step in such a process is to reduce the thermal noise which the oscillator is subject to in order to bring it close to its ground state. 

Typical nanomechanical resonators have frequencies of the order of tens of megahertz. This means that to cool the resonator so that its average energy corresponds to its first excited state requires a temperature of a few milliKelvin. Dilution refrigerators can obtain temperatures of a few hundred milliKelvin, but to reduce the temperature further requires something else. In reference~\cite{Hopkins03} the authors show that, at least in theory, feedback control could be used to obtain the required temperatures. 

\begin{figure}
\centering
\includegraphics[width=2.8in]{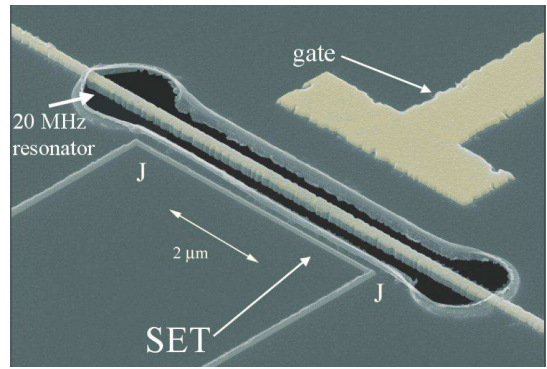} 
\caption{The nanomechanical resonator is a long thin bridge formed on a layered wafer 
by etching out the substrate beneath it. The bridge is driven with a magnetic field, and 
oscillates up and down at $20 \mbox{MHz}$. On the near side of the resonator is a single 
electron transistor (SET), whose central island is formed by the two junctions marked ``J''. 
On the far side is a T-Shaped electrode or ``gate''. The voltage on this gate is varied and 
this results in a varying force on the resonator. Image courtesy of Keith Schwab.}
\label{fig2}
\end{figure}

To perform feedback control one must have a means of monitoring the position of the  resonator and applying a feedback force. The position can be monitored using a single electron transistor, and this has recently been achieved experimentally~\cite{LaHaye04}. A feedback force can be applied by varying the voltage on a gate placed adjacent to the resonator.   This configuration is depicted in Figure~\ref{fig2}. Since the oscillator is harmonic, classical LQG theory can be used to obtain an optimal feedback algorithm for minimizing the energy of the resonator, so long as one takes into account the quantum back-action noise caused by the measurement as described in~\cite{DJ}. The details involved in obtaining the optimal feedback algorithm and calculating the optimal measurement strength are given in~\cite{Hopkins03}. It has further been shown in~\cite{Ruskov05} that adaptive measurement and feedback can be used to prepare the resonator in a squeezed state.

The second example of feedback control we consider is that of cooling an atom trapped in an optical cavity. An optical cavity consists of two parallel mirrors with a single laser beam bouncing back and forward between them. The laser beam forms a standing wave between the mirrors, and if the laser frequency is chosen appropriately, a single atom 
inside the cavity will feel a sinusoidal potential due its interaction with the standing wave. It is therefore possible to trap an atom in one of the wells of this potential. It turns out that information regarding the position of the atom can be obtained by monitoring the phase of the light which leaks out one of the mirrors. Specifically, the phase of the output light tells the observer how far up the side of a potential well the atom is. In addition, by changing the intensity of the laser beam that is driving the cavity, one changes the height of the standing wave, and thus the height of the potential wells. In this system we therefore have a means to monitor the atom and to apply a feedback force. In~\cite{Steck04,Steck06} the authors present a feedback algorithm which can be used to cool the atom to its ground state. 
Actually, the algorithm will prepare the atom either in its ground state, or its first excited state, each with a probability of $50\%$. However, from the measurement record the observer know which one, and can take appropriate action if the resulting state is not the desired one. 

If the location of the atom was known very accurately, then we could use the following feedback algorithm to reduce its energy: increase the height of the potential when the atom is climbing up the side of a well, and reduce it when the atom is falling down towards the centre. This way the energy of the atom is reduced on each oscillation, and the atom will eventually be stationary at the centre of the well. However, it turns out that this algorithm is not effective, either classically or quantum mechanically, when the variance of atom in phase space is appreciable. The reason is that the cyclic process of raising and lowering the potential, which reduces the energy of the atom's mean position and momentum, actually increases the variance of the phase-space probability density. Classically we can eliminate this problem by observing the atom with sufficient accuracy, but quantum mechanically Heisenberg's uncertainty relation prevents us from reducing the variance sufficiently. As a result, an alternative algorithm is required. 

If turns out that one can obtain an effective cooling algorithm by calculating the derivative of the total motional 
energy of the atom with respect to changes in the height of the potential. In doing so one finds that the energy 
change is maximal and minimal at a certain points in the oscillatory motion of the atom. As a result, one can 
use a bang-bang algorithm to switch the potential high when the energy reduction is maximal, and switch 
it low when the resulting energy increase is minimal, in a similar fashion to the classical algorithm described 
above.  The result is that the atom will lose motional energy on each cycle. 

The curious effect whereby the atom will cool to the ground state only half the time is due to the symmetry of the system, and the fact that the feedback algorithm respects this symmetry. Specifically, the feedback process cannot change the average parity of the initial probability density. Since the ground state has even parity, and the first excited state odd parity, if the initial density of the atom has no particular parity (a reasonable assumption), then to preserve this on average the process must pick even and odd final states equally often. Full details regarding the feedback algorithm and the resulting dynamics of the atom is given in~\cite{Steck04,Steck06}.

Although we do not have the space to describe them here, applications of feedback control have been proposed in a variety of other quantum systems. Three of these are cooling the motion of a cavity mirror by modulating the light in the cavity~\cite{Mancini98}, controlling the motion of quantum-dot qubits~\cite{RK}, and preparing spin-squeezed states in atomic clouds~\cite{Thomsen02}. In addition, feedback control has now been experimentally demonstrated in a number of quantum systems. Namely in optics~\cite{Armen02,Smith02}, cold atom clouds~\cite{Geremia04b}, and trapped ions~\cite{Bushev06}.

\section{Conclusion}

To summarize, feedback control has a wide variety of applications in quantum systems, particularly 
in the areas of noise reduction, stabilization, cooling and precision measurement. Such 
applications can be expected to grow more numerous as quantum systems become important 
as the basis of new technologies. While exact results from modern control theory 
can be used to obtain optimal control algorithms for some quantum systems, this is not true for the 
majority of quantum control problems due to their inherent non-linearity. As a result many  
techniques developed for the control of nonlinear classical systems are of considerable 
use in designing algorithms for quantum feedback control, and the efficacy of many of these 
techniques still remain to be explored in quantum systems. Hopefully as further quantum 
control problems arise in specific systems, and effective control algorithms are developed, rules of 
thumb will emerge for the control of classes of quantum devices.

\section*{Acknowledgments}
This work was supported by The Hearne Institute for Theoretical Physics, 
The National Security Agency, The Army Research Office and The Disruptive 
Technologies Office.

\IEEEtriggeratref{39}



\begin{thebibliography}{10}
\providecommand{\url}[1]{#1}
\csname url@rmstyle\endcsname
\providecommand{\newblock}{\relax}
\providecommand{\bibinfo}[2]{#2}
\providecommand\BIBentrySTDinterwordspacing{\spaceskip=0pt\relax}
\providecommand\BIBentryALTinterwordstretchfactor{4}
\providecommand\BIBentryALTinterwordspacing{\spaceskip=\fontdimen2\font plus
\BIBentryALTinterwordstretchfactor\fontdimen3\font minus
  \fontdimen4\font\relax}
\providecommand\BIBforeignlanguage[2]{{%
\expandafter\ifx\csname l@#1\endcsname\relax
\typeout{** WARNING: IEEEtran.bst: No hyphenation pattern has been}%
\typeout{** loaded for the language `#1'. Using the pattern for}%
\typeout{** the default language instead.}%
\else
\language=\csname l@#1\endcsname
\fi
#2}}

\bibitem{Kseq}
P.~S. Maybeck, \emph{Stochastic Models, Estimation and Control}.\hskip 1em plus
  0.5em minus 0.4em\relax Academic Press, New York, 1982, vol. I and II.

\bibitem{Whittle}
P.~Whittle, \emph{Optimal Control}.\hskip 1em plus 0.5em minus 0.4em\relax
  Wiley \& Sons, Chichester, 1996.

\bibitem{WienerIntroPaper}
D.~T. Gillespie, ``The mathematics of brownian motion and johnson noise,''
  \emph{Am. J. Phys.}, vol.~64, p. 225, 1996.

\bibitem{KJPhD}
K.~Jacobs, ``Topics in quantum measurement and quantum noise,'' Ph.D.
  dissertation, Imperial College, London, 1998.

\bibitem{Sakuri}
J.~J. Sakuri, \emph{Modern Quantum Mechanics}.\hskip 1em plus 0.5em minus
  0.4em\relax Addison-Wesley, San Francisco, 1994.

\bibitem{BelavkinLQG}
V.~P. Belavkin, ``Non-demolition measurement and control in quantum dynamical
  systems,'' in \emph{Information, Complexity and Control in Quantum Physics},
  A.~Blaquiere, S.~Diner, and G.~Lochak, Eds.\hskip 1em plus 0.5em minus
  0.4em\relax Springer-Verlag, New York, 1987.

\bibitem{WM93}
H.~M. Wiseman and G.~J. Milburn, ``Quantum theory of field-quadrature
  measurements,'' \emph{Phys. Rev. A}, vol.~47, p. 642, 1993.

\bibitem{Brun02}
T.~A. Brun, ``A simple model of quantum trajectories,'' \emph{American Journal
  of Physics}, vol.~70, p. 719, 2002.

\bibitem{Bouten06}
L.~Bouten, R.~van Handel, and M.~James, ``An introduction to quantum
  filtering,'' \emph{Eprint: math.OC/0601741}.

\bibitem{DJ}
A.~C. Doherty and K.~Jacobs, ``Feedback control of quantum systems using
  continuous state estimation,'' \emph{Phys. Rev. A}, vol.~60, p. 2700, 1999.

\bibitem{DHJMT}
A.~C. Doherty, S.~Habib, K.~Jacobs, H.~Mabuchi, and S.~M. Tan, ``Quantum
  feedback control and classical control theory,'' \emph{Phys. Rev. A},
  vol.~62, p. 012105, 2000.

\bibitem{Yanagisawa98}
M.~Yanagisawa and H.~Kimura, in \emph{Learning, Control and Hybrid Systems,
  Lecture Notes in Control and Information Sciences}.\hskip 1em plus 0.5em
  minus 0.4em\relax Springer-Verlag, New York, 1998, vol. 241, p. 249.

\bibitem{FJ}
C.~A. Fuchs and K.~Jacobs, ``Information-tradeoff relations for finite-strength
  quantum measurements,'' \emph{Phys. Rev. A}, vol.~63, p. 062305, 2001.

\bibitem{DJJ}
A.~C. Doherty, K.~Jacobs, and G.~Jungman, ``Information, disturbance, and
  hamiltonian quantum feedback control,'' \emph{Phys. Rev. A}, vol.~63, p.
  062306, 2001.

\bibitem{DW04}
A.~C. Doherty and H.~M. Wiseman, ``Quantum limits to feedback control of linear
  systems,'' P.~Heszler, Ed., vol. 5468, no.~1.\hskip 1em plus 0.5em minus
  0.4em\relax SPIE, 2004, p. 322.

\bibitem{Yanagisawa03a}
M.~Yanagisawa and H.~Kimura, ``Transfer function approach to quantum
  control-part i: Dynamics of quantum feedback systems,'' \emph{IEEE Trans.
  Automat. Contr.}, vol.~48, p. 2107, 2003.

\bibitem{Yanagisawa03b}
------, ``Transfer function approach to quantum control-part ii: Control
  concepts and applications,'' \emph{IEEE Trans. Automat. Contr.}, vol.~48, p.
  2121, 2003.

\bibitem{DHelon05x}
C.~D'Helon and M.~R. James, ``Stability, gain, and robustness in quantum
  feedback networks,'' Eprint: quant-ph/0511140.

\bibitem{James04}
M.~R. James, ``Risk-sensitive optimal control of quantum systems,'' \emph{Phys.
  Rev. A}, vol.~69, p. 032108, 2004.

\bibitem{Ralph04}
J.~F. Ralph, E.~J. Griffith, T.~D. Clark, and M.~J. Everitt, ``Guidance and
  control in a josephson charge qubit,'' \emph{Phys. Rev. B}, vol.~70, p.
  214521, 2004.

\bibitem{Handel05}
R.~van Handel and H.~Mabuchi, ``Quantum projection filter for a highly
  nonlinear model in cavity qed,'' \emph{J. Opt. B: Quantum Semiclass. Opt.},
  vol.~7, p. S226, 2005.

\bibitem{BEB}
L.~Bouten, S.~Edwards, and V.~P. Belavkin, ``Bellman equations for optimal
  feedback control of qubit states,'' \emph{J. Phys. B: At. Mol. Opt. Phys.},
  vol.~38, p. 151, 2005.

\bibitem{Dolinar73}
S.~Dolinar, \emph{Tech. Rep. 111, Research Laboratory of Electronics}.\hskip
  1em plus 0.5em minus 0.4em\relax MIT, Cambridge, 1973.

\bibitem{Geremia04}
J.~Geremia, ``Distinguishing between optical coherent states with imperfect
  detection,'' \emph{Phys. Rev. A}, vol.~70, p. 062303, 2004.

\bibitem{Helstrom}
C.~W. Helstrom, \emph{Quantum Detection and Estimation Theory}, ser.
  Mathematics in Science and Egineering.\hskip 1em plus 0.5em minus 0.4em\relax
  Academic Press, New York, 1976, vol. 123.

\bibitem{Jacobs06x}
K.~Jacobs, ``Feedback control for communication with non-orthogonal states,''
  \emph{Eprint: quant-ph/0601162}.

\bibitem{Milburn94}
G.~J. Milburn, K.~Jacobs, and D.~F. Walls, ``Quantum-limited measurements with
  the atomic force microscope,'' \emph{Phys. Rev. A}, vol.~50, no.~6, p. 5256,
  1994.

\bibitem{Braginsky03}
V.~B. Braginsky, M.~L. Gorodetsky, F.~Y. Khalili, A.~B. Matsko, K.~S. Thorne,
  and S.~P. Vyatchanin, ``Noise in gravitational-wave detectors and other
  classical-force measurements is not influenced by test-mass quantization,''
  \emph{Phys. Rev. D}, vol.~67, no.~8, p. 082001, 2003.

\bibitem{Verstraete01}
F.~Verstraete, A.~C. Doherty, and H.~Mabuchi, ``Sensitivity optimization in
  quantum parameter estimation,'' \emph{Phys. Rev. A}, vol.~64, p. 032111,
  2001.

\bibitem{Stockton04}
J.~K. Stockton, J.~M. Geremia, A.~C. Doherty, and H.~Mabuchi, ``Robust quantum
  parameter estimation: Coherent magnetometry with feedback,'' \emph{Phys. Rev.
  A}, vol.~69, p. 032109, 2004.

\bibitem{Geremia05}
J.~Geremia, J.~K. Stockton, and H.~Mabuchi, ``Suppression of spin projection
  noise in broadband atomic magnetometry,'' \emph{Physical Review Letters},
  vol.~94, p. 203002, 2005.

\bibitem{Berry01}
D.~W. Berry, H.~M. Wiseman, and J.~K. Breslin, ``Optimal input states and
  feedback for interferometric phase estimation,'' \emph{Phys. Rev. A},
  vol.~63, p. 053804, 2001.

\bibitem{Berry02}
D.~W. Berry and H.~M. Wiseman, ``Adaptive quantum measurements of a
  continuously varying phase,'' \emph{Physical Review A}, vol.~65, p. 043803,
  2002.

\bibitem{Pope04}
D.~T. Pope, H.~M. Wiseman, and N.~K. Langford, ``Adaptive phase estimation is
  more accurate than nonadaptive phase estimation for continuous beams of
  light,'' \emph{Physical Review A}, vol.~70, p. 043812, 2004.

\bibitem{Pegg88}
D.~T. Pegg and S.~M. Barnett, ``Unitary phase operator in quantum mechanics,''
  \emph{Europhysics Lett.}, vol.~6, p. 483, 1988.

\bibitem{Pregnell06}
K.~L. Pregnell and D.~T. Pegg, ``Single-shot measurement of quantum optical
  phase,'' \emph{Eprint: quant-ph/0506206}.

\bibitem{Wiseman95}
H.~M. Wiseman, ``Adaptive phase measurements of optical modes: Going beyond the
  marginal q distribution,'' \emph{Phys. Rev. Lett.}, vol.~75, p. 4587, 1995.

\bibitem{Wiseman97}
H.~M. Wiseman and R.~B. Killip, ``Adaptive single-shot phase measurements: A
  semiclassical approach,'' \emph{Phys. Rev. A}, vol.~56, p. 944, 1997.

\bibitem{Wiseman98}
------, ``Adaptive single-shot phase measurements: The full quantum theory,''
  \emph{Phys. Rev. A}, vol.~57, p. 2169, 1998.

\bibitem{Armen02}
M.~A. Armen, J.~K. Au, J.~K. Stockton, A.~C. Doherty, and H.~Mabuchi,
  ``Adaptive homodyne measurement of optical phase,'' \emph{Phys. Rev. Lett.},
  vol.~89, p. 133602, 2002.

\bibitem{Spiller05}
T.~P. Spiller, W.~J. Munro, S.~D. Barrett, and P.~Kok, ``An introduction to
  quantum information processing: applications and realizations,'' \emph{Cont.
  Phys.}, vol.~46, p. 407, 2005.

\bibitem{rapidP}
K.~Jacobs, ``How to project qubits faster using quantum feedback,'' \emph{Phys.
  Rev. A}, vol.~67, p. 030301(R), 2003.

\bibitem{KJSPIE04}
------, ``Optimal feedback control for rapid preparation of a qubit,''
  P.~Heszler, Ed., vol. 5468, no.~1.\hskip 1em plus 0.5em minus 0.4em\relax
  SPIE, 2004, p. 355.

\bibitem{Wiseman06x}
H.~M. Wiseman and J.~F. Ralph, ``Reconsidering rapid qubit purification by
  feedback,'' \emph{Eprint: quant-ph/0603062}.

\bibitem{Ralph06}
J.~F. Ralph, E.~J. Griffith, and C.~D. H. A.~D. Clark, ``Rapid purification of
  a solid state charge qubit,'' E.~J. Donkor, A.~R. Pirich, and H.~E. Brandt,
  Eds., vol. 6244.\hskip 1em plus 0.5em minus 0.4em\relax SPIE, 2006, p. (in
  press).

\bibitem{Griffith06x}
E.~J. Griffith, C.~D. Hill, J.~F. Ralph, and K.~Jacobs, ``Rapid state
  purification in a superconducting charge qubit,'' \emph{(in preparation)}.

\bibitem{Combes06}
J.~Combes and K.~Jacobs, ``Rapid state-reduction of quantum systems using
  feedback control,'' \emph{Phys. Rev. Lett.}, vol.~96, p. 010504, 2006.

\bibitem{SHM}
J.~K. Stockton, R.~van Handel, and H.~Mabuchi, ``Deterministic dicke-state
  preparation with continuous measurement and control,'' \emph{Phys. Rev. A},
  vol.~70, p. 022106, 2004.

\bibitem{vanHandel05}
R.~van Handel, J.~K. Stockton, and H.~Mabuchi, ``Feedback control of quantum
  state reduction,'' \emph{IEEE Trans. Automat. Control}, vol.~50, p. 768,
  2005.

\bibitem{Ahn02}
C.~Ahn, A.~C. Doherty, and A.~J. Landahl, ``Continuous quantum error correction
  via quantum feedback control,'' \emph{Physical Review A (Atomic, Molecular,
  and Optical Physics)}, vol.~65, p. 042301, 2002.

\bibitem{Sarovar04}
M.~Sarovar, C.~Ahn, K.~Jacobs, and G.~J. Milburn, ``Practical scheme for error
  control using feedback,'' \emph{Physical Review A (Atomic, Molecular, and
  Optical Physics)}, vol.~69, p. 052324, 2004.

\bibitem{Ahn04}
C.~Ahn, H.~Wiseman, and K.~Jacobs, ``Quantum error correction for continuously
  detected errors with any number of error channels per qubit,'' \emph{Phys.
  Rev. A}, vol.~70, p. 024302, 2004.

\bibitem{Blencowe04}
M.~P. Blencowe, ``Quantum electromechanical systems,'' \emph{Phys. Rep.}, vol.
  395, p. 159, 2004.

\bibitem{Cho03}
A.~Cho, ``Researchers race to put the quantum into mechanics,'' \emph{Science},
  vol. 299, no. 5603, p.~36, 3 January 2003.

\bibitem{Hopkins03}
A.~Hopkins, K.~Jacobs, S.~Habib, and K.~Schwab, ``Feedback cooling of a
  nanomechanical resonator,'' \emph{Phys. Rev. B}, vol.~68, p. 235328, 2003.

\bibitem{LaHaye04}
M.~D. LaHaye, O.~Buu, B.~Camarota, and K.~C. Schwab, ``Approaching the quantum
  limit of a nanomechanical resonator,'' \emph{Science}, vol. 304, p.~74, 2004.

\bibitem{Ruskov05}
R.~Ruskov, K.~Schwab, and A.~N. Korotkov, ``Squeezing of a nanomechanical
  resonator by quantum nondemolition measurement and feedback,'' \emph{Phys.
  Rev. B}, vol.~71, p. 235407, 2005.

\bibitem{Steck04}
D.~Steck, K.~Jacobs, H.~Mabuchi, T.~Bhattacharya, and S.~Habib, ``Quantum
  feedback control of atomic motion in an optical cavity,'' \emph{Phys. Rev.
  Lett.}, vol.~92, p. 223004, 2004.

\bibitem{Steck06}
D.~Steck, K.~Jacobs, H.~Mabuchi, S.~Habib, and T.~Bhattacharya, ``Feedback
  cooling of atomic motion in cavity qed,'' \emph{Eprint: quant-ph/0509039}.

\bibitem{Mancini98}
S.~Mancini, D.~Vitali, and P.~Tombesi, ``Optomechanical cooling of a
  macroscopic oscillator by homodyne feedback,'' \emph{Phys. Rev. Lett.},
  vol.~80, p. 688, 1998.

\bibitem{RK}
R.~Ruskov and A.~N. Korotkov, ``Quantum feedback control of a solid-state
  qubit,'' \emph{Phys. Rev. B}, vol.~66, p. 041401, 2002.

\bibitem{Thomsen02}
L.~K. Thomsen, S.~Mancini, and H.~M. Wiseman, ``Spin squeezing via quantum
  feedback,'' \emph{Phys. Rev. A}, vol.~65, p. 061801, 2002.

\bibitem{Smith02}
W.~P. Smith, J.~E. Reiner, L.~A. Orozco, S.~Kuhr, and H.~M. Wiseman, ``Capture
  and release of a conditional state of a cavity qed system by quantum
  feedback,'' \emph{Phys. Rev. Lett.}, vol.~89, p. 133601, 2002.

\bibitem{Geremia04b}
J.~M. Geremia, J.~K. Stockton, and H.~Mabuchi, ``Real-time quantum feedback
  control of atomic spin-squeezing,'' \emph{Science}, vol. 304, p. 270, 2004.

\bibitem{Bushev06}
P.~Bushev, D.~Rotter, A.~Wilson, F.~Dubin, C.~Becher, J.~Eschner, R.~Blatt,
  V.~Steixner, P.~Rabl, and P.~Zoller, ``Feedback cooling of a single trapped
  ion,'' \emph{Phys. Rev. Lett.}, vol.~96, p. 043003, 2006.

\end{thebibliography}

\end{document}